\documentclass[times,twocolumn,final,authoryear]{elsarticle}

\usepackage{jasr}
\usepackage{framed,multirow}

\usepackage{amssymb}
\usepackage{latexsym}

\usepackage[switch]{lineno}

\usepackage{url}
\usepackage{xcolor}
\definecolor{newcolor}{rgb}{.8,.349,.1}

\usepackage[citebordercolor=white]{hyperref}

\journal{Advances in Space Research}

\newcommand{\arcsec}{$^\prime$$^\prime$}
\newcommand{\degr}{$^\circ$}

\begin{document}

\verso{Given-name Surname \textit{etal}}

\begin{frontmatter}

\title{Where is the base of the Transition Region? Evidence from TRACE, SDO, IRIS and ALMA observations \tnoteref{tnote1}}%

\author[1]{Costas E. \snm{Alissandrakis}\corref{cor1}}
\cortext[cor1]{Corresponding author: 
  Tel.: +30-26510-08492;  
  fax: +0-000-000-0000;}
\ead{calissan@uoi.gr}

\address[1]{Department of Physics, University of Ioannina, GR45110 Ioannina, Greece}

\received{}
\finalform{}
\accepted{}
\availableonline{}
\communicated{}

\begin{abstract}
{ Classic} solar atmospheric models { put} the Chromosphere-Corona Transition Region ({ CCTR}) at $\sim2$\,Mm above the $\tau_{5000} = 1$ level{, whereas radiative MHD (rMHD) models place the CCTR in a wider range of heights.} However, observational verification is scarce. In this work we review and discuss recent results from various instruments and spectral domains. In SDO and TRACE images spicules appear in emission in the 1600, 1700 and 304\,\AA\ bands and in absorption in the EUV bands; the latter is due to photo-ionization of H{\sc\,i} and He{\sc\,i}, which increases with wavelength. At the shortest available AIA wavelength and taking into account that the photospheric limb is  $\sim0.34$\,Mm above the $\tau_{5000} = 1$ level, we found that { CCTR} emission starts at $\sim3.7$\,Mm; extrapolating to $\lambda=0$, where there is no chromospheric absorption, we deduced a height of $3.0 \pm 0.5$\,Mm, which is above the value of 2.14\,Mm of the Avrett and Loeser model. Another indicator of the extent of the chromosphere is the height of the network structures. Height differences produce a limbward shift of features with respect to the position of their counterparts in magnetograms. Using this approach, we measured heights of $0.14 \pm 0.04$\,Mm (at 1700\,\AA), $0.31 \pm 0.09$\,Mm (at 1600\,\AA) and $3.31 \pm 0.18$\,Mm (at 304\,\AA) for the center of the solar disk. A previously reported possible solar cycle variation is not confirmed. A third indicator is the position of the limb in the UV, where IRIS observations of the Mg{\sc\,ii} triplet lines show that they extend up to $\sim2.1$\,Mm above the 2832\,\AA\ limb, while AIA/SDO images give a limb height of $1.4 \pm 0.2$\,Mm (1600\,\AA) and $5.7 \pm 0.2$\,Mm (304\,\AA). Finally, ALMA mm-$\lambda$ full-disk images provide useful diagnostics, though not very accurate, due to their relatively low resolution; values of $2.4 \pm 0.7$\,Mm at 1.26\,mm and $4.2 \pm 2.5$\,Mm at 3\,mm were obtained. Putting everything together, we conclude that the average chromosphere extends higher than homogeneous models predict{, but within the range of rMHD models.}.
\end{abstract}

\begin{keyword}
\KWD Sun\sep Chromosphere-corona transition region\sep Solar atmospheric models
\end{keyword}

\end{frontmatter}


\section{Introduction} \label{intro} 
The direct measurement of the height of the solar emission in various spectral regions is very important for the computation and verification of atmospheric models. In physical terms, the formation height of the radiation is determined by the local conditions and radiative transfer. We note that in classic empirical atmospheric models the height is not a basic parameter, but it is derived from the optical depth and the absorption coefficient (see, e.g., Zirin, 1966).). The formation height of the emission at a particular wavelength is computed from { the emissivity and opacity, which depend on the atmospheric model}
(see, e.g. Figure 1 of Vernazza et al., 1976). 

\begin{figure*}[t]
  \centering
  \includegraphics[height=6cm, width=5cm]{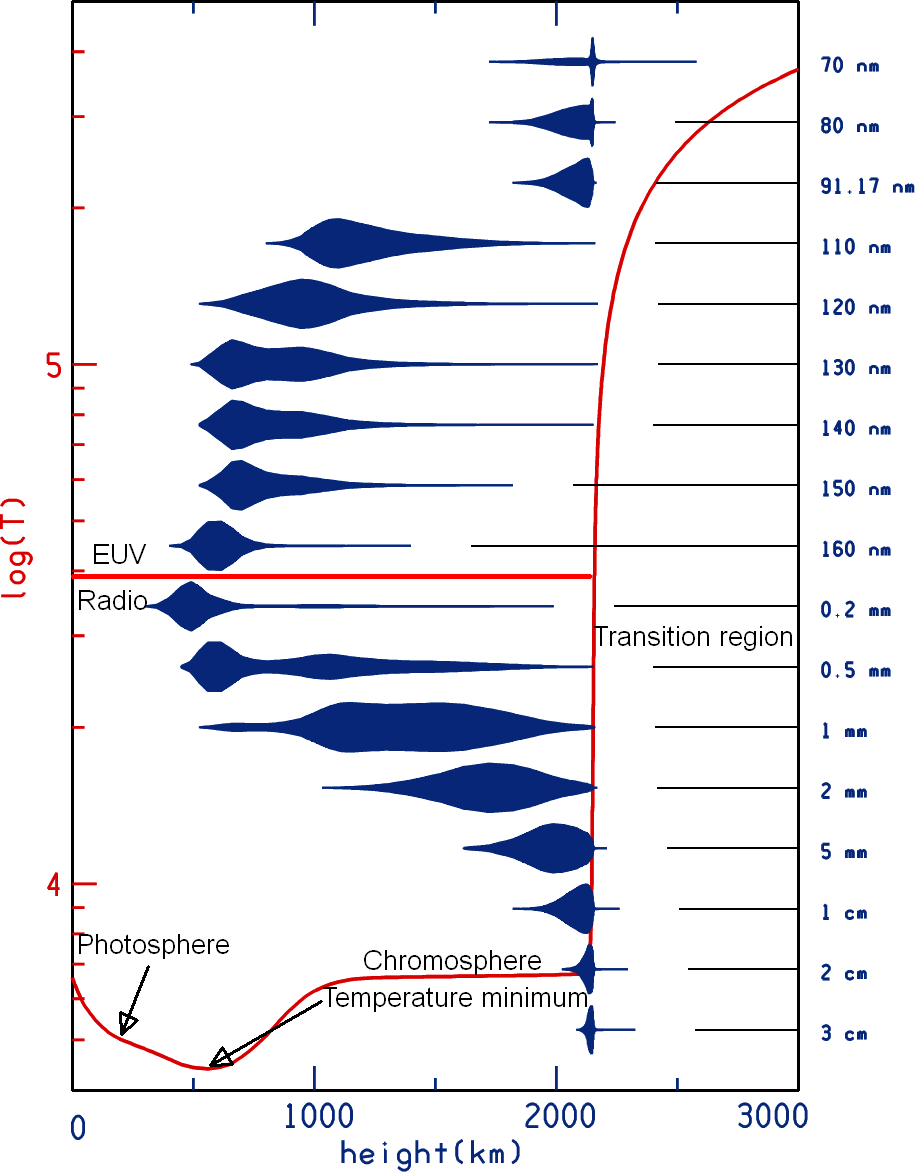}\hspace{.5cm}\includegraphics[height=6cm]{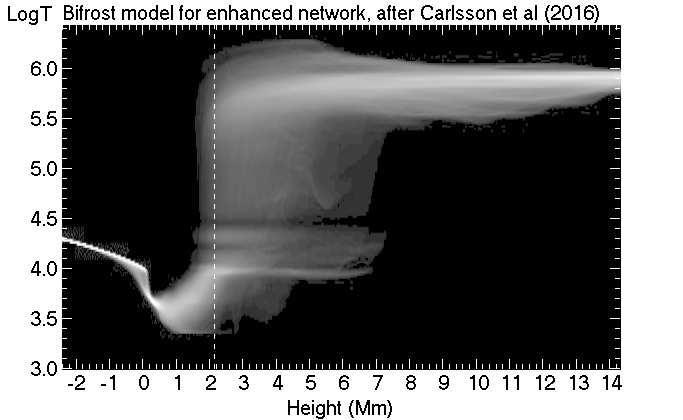}
  \caption{{ Left: The electron temperature as a function of height (red line) according to model C7 of Avrett \& Loeser, 2008. The contribution to the continuum intensity is plotted in blue for the short radio wavelengths (below the red horizontal line) and the extreme ultraviolet wavelengths, marked at the right. Adapted from Fig. 2 of Avrett \& Loeser, 2008.. Right: Probability density function of the temperature as function of height from Bifrost simulations of an enhanced network region. The vertical dashed line is at 2.14\,Mm. Adapted from the lower panel of Fig. 10 of Carlsson et al., 2016.}}
  \label{fig:models}
\end{figure*}

{ Classical semi-empirical models assume that the solar atmosphere is horizontally homogeneous}
and in hydrostatic equilibrium, 
{ with an
additional {\it ad hoc} turbulent pressure to set the extent of the chromosphere; they ignore mass motions and the associated dynamic pressure, as well as Lorentz forces which produce}
the rich horizontal structure that is more prominent in the upper layers. Multi-component models take into the account differences between the network and internetwork regions, but still ignore the radiative transfer in the horizontal direction. Nevertheless, they represent a reasonable approximation of the one-dimensional structure of the solar atmosphere. Such models put the base of the Chromosphere-Corona Transition Region ({ CCTR}) at a height of about 2\,000\,km ({\it i.e.} a mere 3\arcsec) above the $\tau_{\rm5000}=1$ level ({\it e.g.} Vernazza et al., 1981; Fontenla et al., 1990, 1991, 1993; Avrett \& Loeser, 2008; Avrett et al., 2015; { also left panel of Fig.~\ref{fig:models}}). This is well below the top of spicules, which protrude from the chromosphere into the corona, to a height that may exceed 15\arcsec\ ({\it e.g.} Alissandrakis et al., 2018). 

Thanks to the development of fast numerical computations, a number of sophisticated tools, such as the Bifrost radiative magnetohydrodynamics (rMHD) code (Gudiksen et al., 2011) and the STockholm inversion Code (STiC, de la Cruz Rodr\'iguez et al., 2019) have been developed for solar atmospheric modeling. { Such models demonstrate well the complex structure and the dynamics of the upper layers of the solar atmosphere. An example of the deduced temperature structure is given in the right panel of Fig.~\ref{fig:models}, from a simulation of an enhanced network region by Carlsson et al. (2016). It is interesting that transition region temperatures, from various features, appear in an extended range of heights that go beyond 2.14\,Mm (dashed line). Although this particular model may not be representative of the average QS, which includes inter-network regions, it is still indicative.} Nevertheless, the classic models still provide a 
{ reference} picture of the solar atmosphere.

Observational evidence for the position of the base of the { CCTR} came originally from eclipse data, and is based on the relative timing between the instance that the lunar disk covers the photosphere and the instance at which coronal lines appear after the occultation of the chromosphere by the moon (Zirin, 1966). Subsequently, Avrett \& Loeser (2008) placed the top of the chromosphere at 2140\,km, roughly consistent with the maximum intensity of the 10833\,\AA\ He{\sc\,i} and the the 5875\,\AA\ D$_3$ lines.

The observable quantity most directly associated with the height of formation is the solar radius. Such measurements were compiled by Rozelot et al. (2015) for the optical and extreme ultraviolet (EUV) and by by Menezes \& Valio (2017) for the microwave range (see also Battaglia et al., 2017 for a measurement in X-rays). We note that direct measurements beyond the limb in the EUV and the X-rays are difficult, due to the low spatial resolution of the older space-born instruments. Daw et al. (1995) used the {\it Normal Incidence X-ray Telescope} (NIXT) to measure the position of the limb at 63.5\,\AA\ with a spatial resolution of $\sim1$\arcsec, while Zhang et al. (1998) used the {\it Extreme-Ultraviolet Imaging Telescope} (EIT) aboard the {\it Solar and Heliospheric Observatory}, (SoHO) at 171, 195, 284 and 304\,\AA\ with $\sim3$\arcsec\ resolution. 
 
Although both the {\it Transition Region and Coronal Explorer} (TRACE) and the {\it Solar Dynamics Observatory} (SDO) have sufficient resolution (of the order of 1\arcsec), the measurement requires precise relative pointing among white light (WL) observations and observations in the EUV channels. In a recent work (Alissandrakis \& Valentino, 2019b, hereafter Paper I; see also Alissandrakis \& Valentino, 2019a for a correction), we used the transits of Mercury and Venus to obtain accurate pointing information and, subsequently, we measured the height of the limb with respect to the photosphere from TRACE and {\it Atmospheric Imaging Assembly} (AIA) images. The pointing problem does not exist in the case of the {\it Interface Region Imaging Spectrograph} (IRIS),
{ where the position of the photospheric limb can be measured directly in the far wing of the k line,} and Alissandrakis et al. (2018) gave values of the limb height for a number of spectral lines and continua with respect to the limb at 2832\,\AA.

On the solar disk, height differences translate to shifts between features observed at different wavelengths. Such measurements were performed by Alissandrakis (2019), hereafter paper II, where the height of network structures seen in AIA images at 1600\,\AA, 1700\,\AA\ and 304\.\AA\ with respect to the associated magnetic structures HMI images was computed.

\begin{figure*}[t]
  \centering
  \includegraphics[width=15cm]{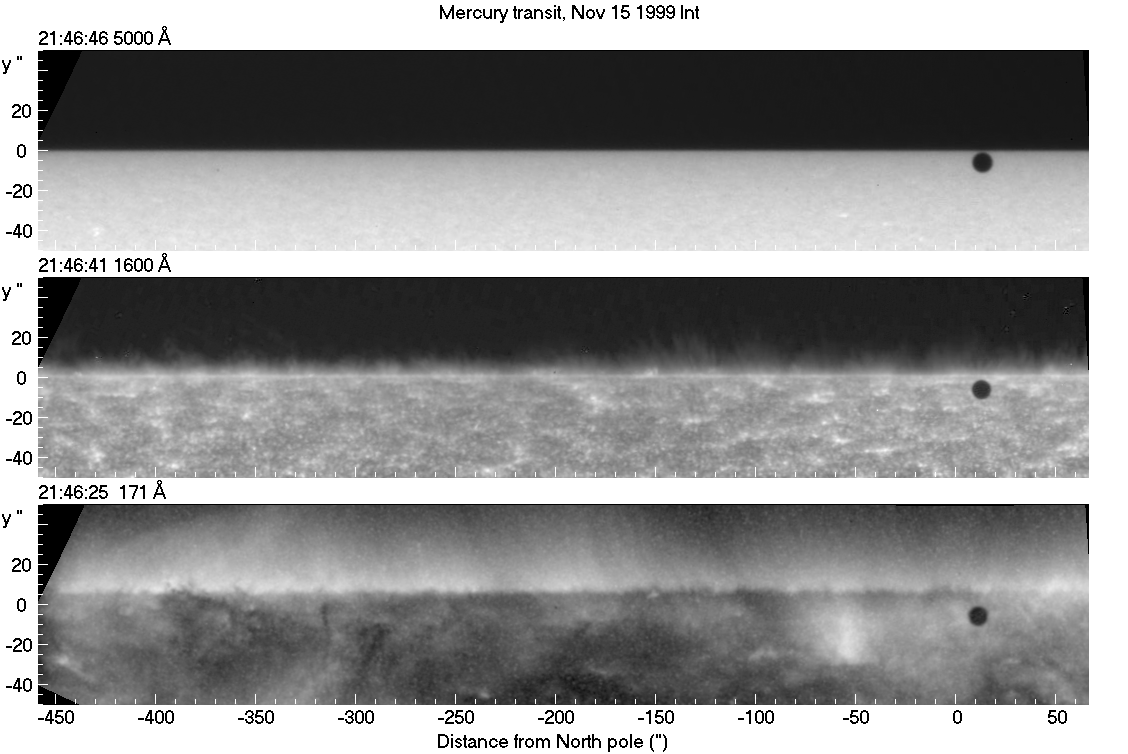}
  \caption{TRACE images near the limb in polar projection, in white light (top), the 1600\,\AA\ band and the 171\,\AA\ band, after pointing correction. The dark disk is Mercury. From Alissandrakis \& Valentino (2019b), reprinted by permission from: Springer Nature, Solar Physics, \copyright 2019.}
  \label{fig:01}
\end{figure*}

At the other side of the spectrum, microwave and mm-$\lambda$ observations near the limb give important information on the { CCTR} height and structure. In this range the opacity, and hence the limb height, increases with wavelength, but the situation is complicated by the low resolution of full-disk observations and the presence of spicules. A compilation of results from {\it Atacama Large Millimeter/submillimeter Array} (ALMA) full-disk images and from AIA UV bands was given in Table 4 of  Alissandrakis et al. (2020). According to that, mm-$\lambda$ radiation forms between the levels of 1600\,\AA\ and 304\,\AA\ emission, while limb heights  of $2.4 \pm 0.7$\,Mm and $4.2 \pm 2.5$\,Mm were reported at 1.26\,mm and 3\,mm respectively.

An important point to note is that the white light limb is above the $\tau_{5000}=1$ level, which is the reference point for zero height. Athay (1976) gave a height difference of 340\,km in his Table I-1, whereas Thuillier et al. (2011) gave similar values: 314 to 346\,km at 4000\,\AA, depending on the model used (their Table 5), with a slight increase with wavelength. Consequently, a correction must be applied to the measured values in order to compare them with model predictions.

In Section 2 of this work we present a brief summary of our previous results. Subsequently, in Sect.~\ref{cycle} we investigate the variation of the network height during the solar cycle and in Sect.~\ref{discuss} we discuss the combined results. 

\begin{figure}
  \centering
  \includegraphics[width=7cm]{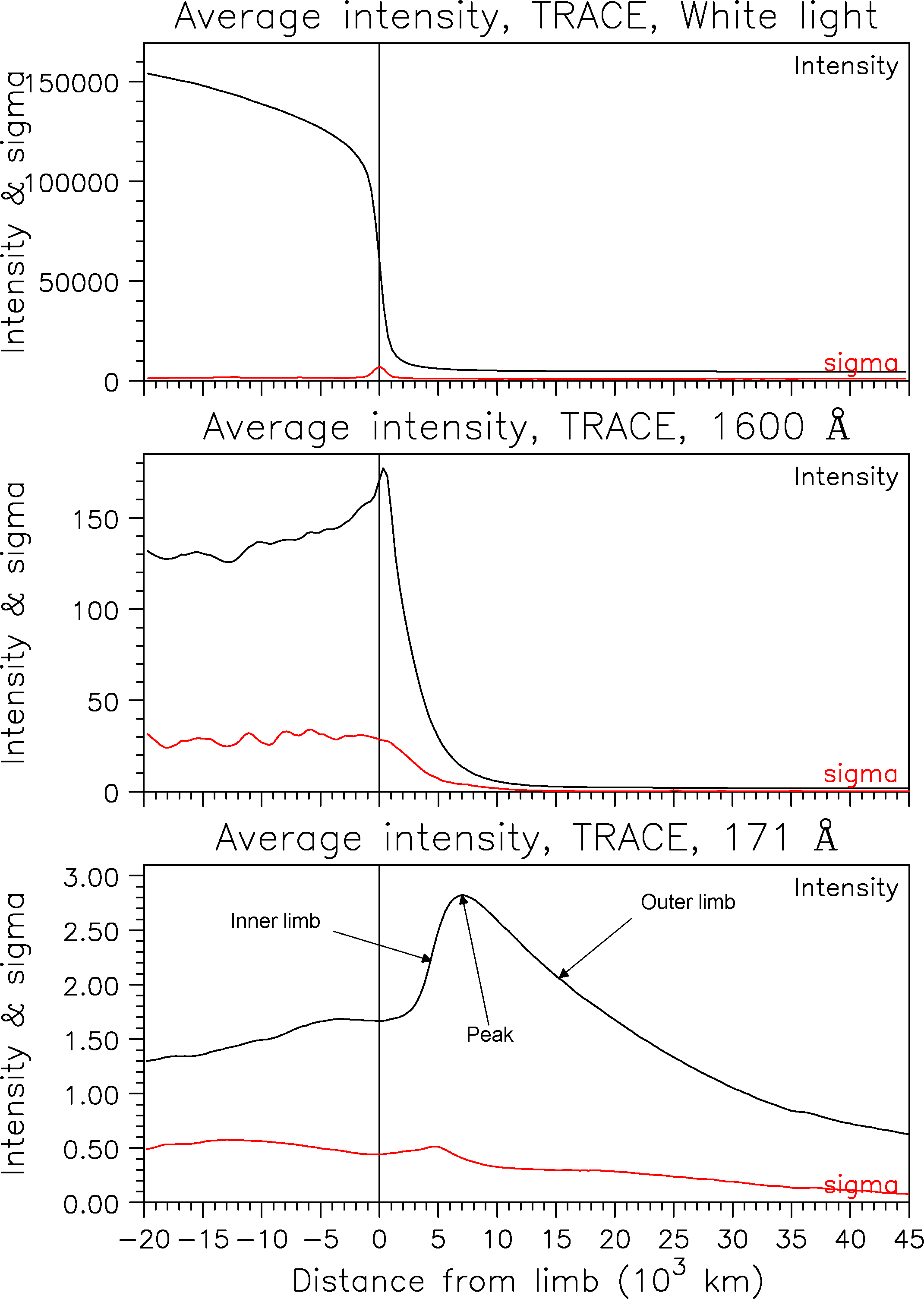}
  \caption{The azimuthally averaged intensity, as a function of distance from the white-light limb for the
three TRACE wavelength bands shown in Fig.~\ref{fig:01}. Curves in red show the rms of the respective average values, which is a measure of the azimuthal intensity variation at each radial position. Adapted from Alissandrakis \& Valentino (2019b).}
  \label{fig:02}
\end{figure}

\section{Review of previous results}

\subsection{Measurements of the location of the { CCTR}}
Beyond the photospheric limb, spicules are well visible in the TRACE UV images (Alissandrakis et al., 2005) and in the SDO 1600\,\AA, 1700\,\AA\ and 304\,\AA\ band images; they are also seen as absorbing features in the EUV images of both instruments, where they give a rugged appearance to the lower boundary of the corona near the limb. This is demonstrated in Fig.~\ref{fig:01}, which shows TRACE images near the limb, corrected for pointing using the position of Mercury, in the white light (WL), 1600 and 171\,\AA\ bands.

Fig.\ref{fig:02} shows plots of the azimuthally averaged intensity for the three wavelength bands displayed in Fig.~\ref{fig:01}. The position of the photospheric limb, used as a reference for zero height, can be easily measured from the inflection point of the corresponding curve. In the 1600\,\AA\ emission, which comes from the { lower} chromosphere { around the temperature minimum}, the curve shows limb brightening that peaks slightly above the photospheric limb, whereas the 171\,\AA\ emission, formed in the { CCTR}, the peak is about 6\,Mm above the WL limb. The inflection point of the intensity curve before the peak defines the {\it inner limb}, and corresponds to the height where the spicule forest becomes transparent, so that emission form the back side of the plane of the sky can reach the observer.

\begin{figure}[t]
  \centering
  \includegraphics[width=8.8cm]{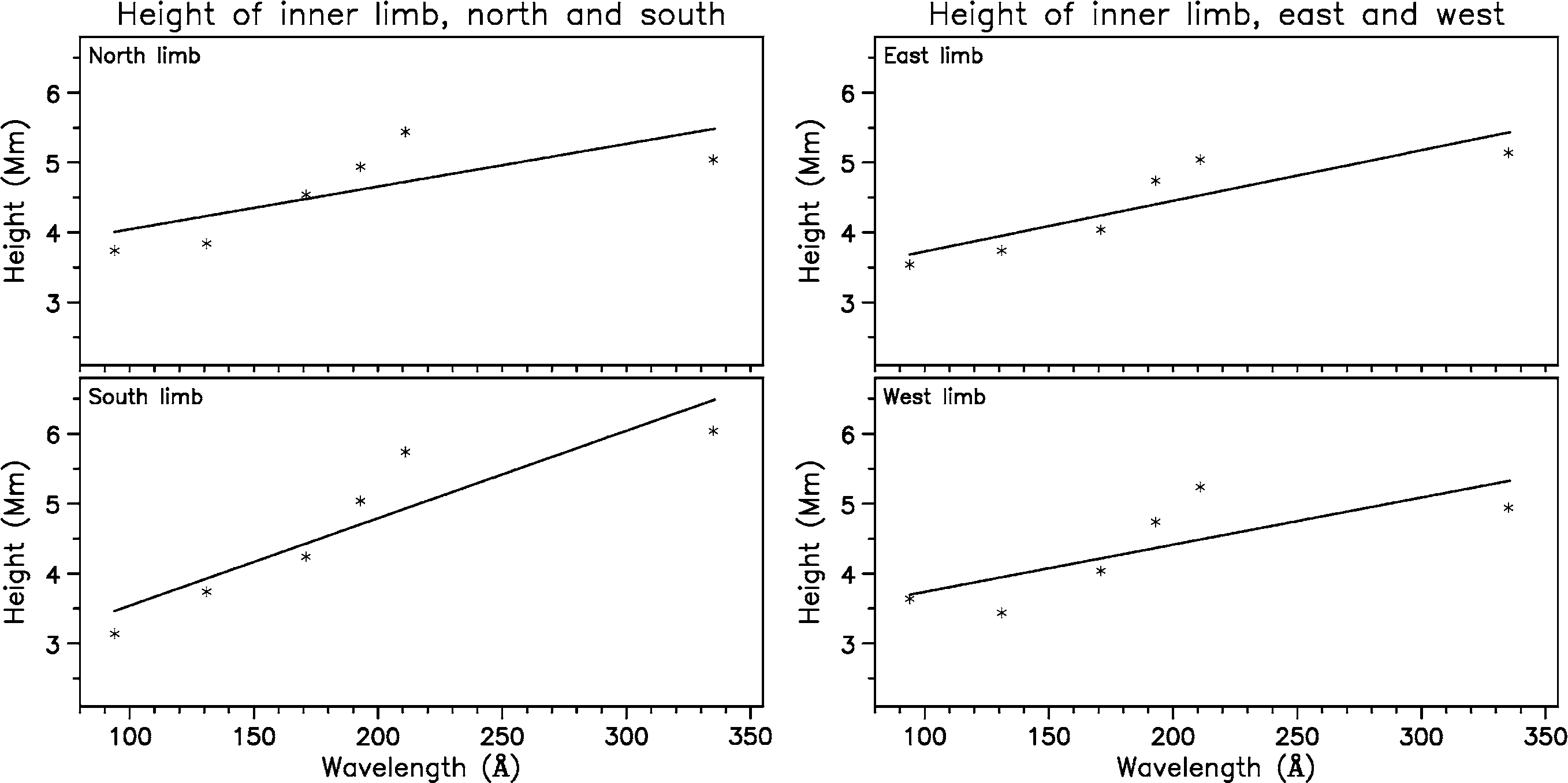}
  \caption{The height of the inner limb with respect to the $\tau_{5000} = 1$ level for the high temperature AIA channels as a function of wavelength. The straight lines show the result of a linear regression. From Alissandrakis \& Valentino (2019a), reprinted by permission from: Springer Nature, Solar Physics, \copyright 2019.}
  \label{fig:03}
\end{figure}

Measurements of the inner limb position in all AIA high-temperature bands (94, 131, 171, 193, 211 and 335\,\AA) were performed in Paper I, after small pointing and image scale corrections derived from the position of Venus during its transit in June 5-6, 2012. The height of the inner limb measured from the $\tau_{5000} = 1$ level is plotted as a function of wavelength  in Fig.~\ref{fig:03}, for regions around the north, south, east and west limbs. 

The position of the inner limb places an upper limit to the height of the base of the { CCTR}, as the presence of spicules does not allow a direct measurement of the average height of the top of the homogeneous chromosphere. This upper limit, deduced from the minimum height of the inner limb at 94\,\AA, is about 3.7\,Mm above the $\tau_{5000} = 1$ level. 

Spicule absorption at wavelengths shorter than the { hydrogen Lyman-continuum head}
is attributed to photoionization, principally of neutral hydrogen (H{\sc\,i}) and neutral or singly-ionized helium (He{\sc\,i} and He{\sc\,ii}), with the absorption coefficient increasing with wavelength; this is in conformity with our results of Fig.~\ref{fig:03}, which show that the height of the inner limb increases with wavelength. Therefore, a better estimate of the height of the { CCTR} can be obtained from the extrapolation of the regression line of the average inner limb height versus wavelength plot (Fig.~\ref{fig:03}) to 
{ to low $\lambda$ values, where there is negligible chromospheric absorption.}
This gives a height of ($3.0 \pm 0.5$)\,Mm, { where the height of the WL limb with respect to the $\tau_{5000} = 1$ level has been taken into account; this} range is above the value of 2.140\,Mm used by Avrett \& Loeser (2008).

{ We note that the linear fit of the variation of height with wavelength in Fig.~\ref{fig:03} is a reasonable approximation, as a full computation of the height as a function of wavelength requires knowledge of H and He densities and degrees of ionization that we do not have. We further note that 212\,\AA\ is below the ionization limit of He{\sc\,ii}, and that the greater value of height in this band could be attributed to the associated opacity jump; however, our computations showed that the opacity jump is small, 6\% to 15\%, for reasonable degrees of He ionization of 10\% and 20\% respectively (see also Table 3 in Anzer \& Heinzel, 2005).}

\subsection{Measurements of the height of network structures}
In Paper II, we measured the shift of network structures in AIA 1600, 1700 and 304\,\AA\ images with respect to the corresponding features in HMI magnetograms, at several distances from the disc center; the shift was measured by cross-correlation, after dividing each image into 12 sectors, 30\degr\ wide; each sector was further divided into a number of zones, 60\arcsec\ wide, from just inside the limb down to a distance of about 0.25 solar radii from the center of the disk. The absolute value of the longitudinal magnetic field was used for the cross-correlation. 
 
\begin{figure}[h]
  \centering
  \includegraphics[width=8cm]{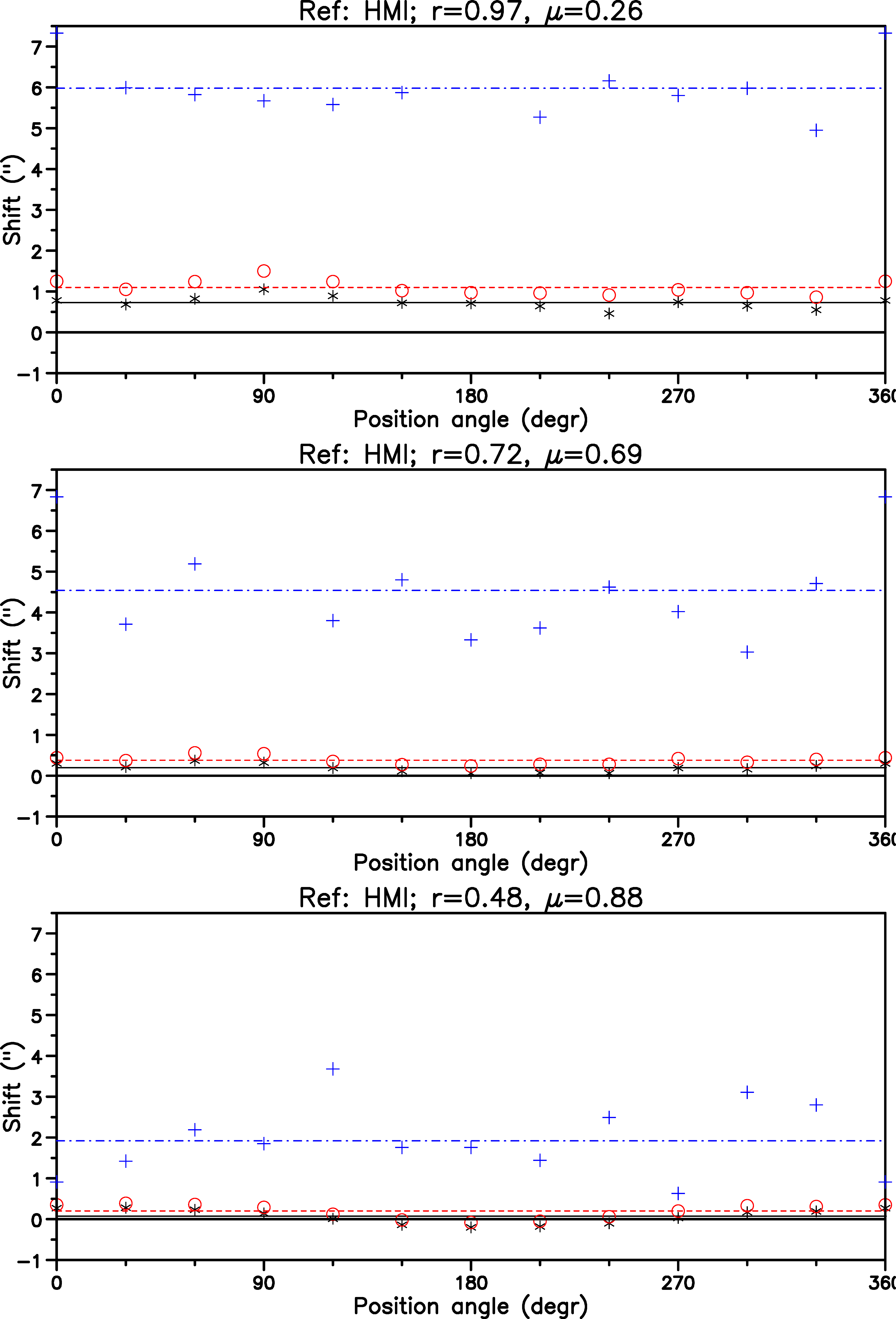}
  \caption{The radial shift of network structures with respect to the corresponding magnetic features as a function of position angle for three locations on the solar disk. The  distance from the center of the disk in solar radii, $r$, and the corresponding cosine of the heliocentric angle, $\mu$, are marked above each panel. The horizontal lines mark the shift averaged over the position angle. Black asterisks and full lines are for 1700 Å, red circles and dotted lines for 1600 Å, and blue crosses and dash-dotted lines for 304 Å. Adapted from paper II.}
  \label{fig:04}
\end{figure}

Representative results are shown in Fig.~\ref{fig:04}. There is no systematic variation with position angle, which points against any pole-equator differences; thus the scatter of the data points reflects measurement errors. It is obvious that the emission in the AIA 304\,\AA\ band forms well above the that in the 1600\,\AA\ band, and that the latter forms slightly above the 1700\,\AA\ band. As expected, the shifts are larger near the limb. Similar results were obtained by Rutten (2020).

\begin{figure}
  \centering
  \includegraphics[width=8.8cm]{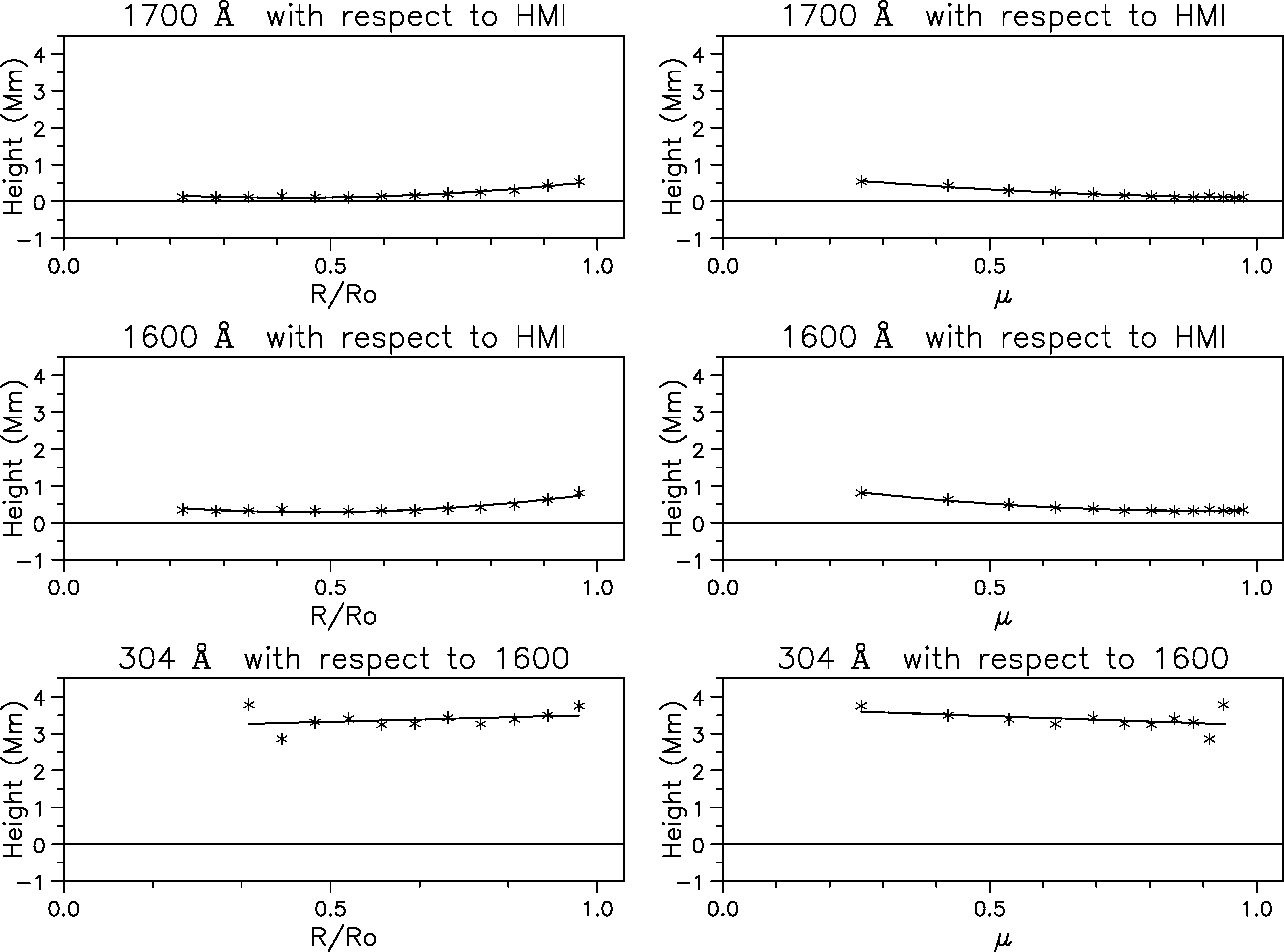}
  \caption{The height of network emission, plotted as a function of distance from the center of the disk (left column) and as a function of the cosine of the heliocentric angle, $\mu$. The reference wavelength and the image wavelength are marked above each panel. Solid lines show the result of linear regression for 304 Å and a quadratic regression for the other wavelengths. Adapted from paper II.}
  \label{fig:05}
\end{figure}

The network height, derived from the shifts, is plotted in Fig.~\ref{fig:05}, as a function of distance from the center of the disk (left column) and as a function of the cosine of the heliocentric angle, $\mu$, (right column). The plots show a well-defined increase of the network height from the center of the disk to the limb, in all three wavelength bands. This is apparently due to opacity effects, as the radiation near the limb is formed higher than near the disk center. The lowest height is that of the 1700 Å network, just 0.14 Mm above the HMI continuum level at the disk center, whereas the 304 Å network height is 3.3 Mm, again  at the disk center.

\subsection{Limb position from IRIS data}
\begin{figure}[h]
  \centering
  \includegraphics[width=8.8cm]{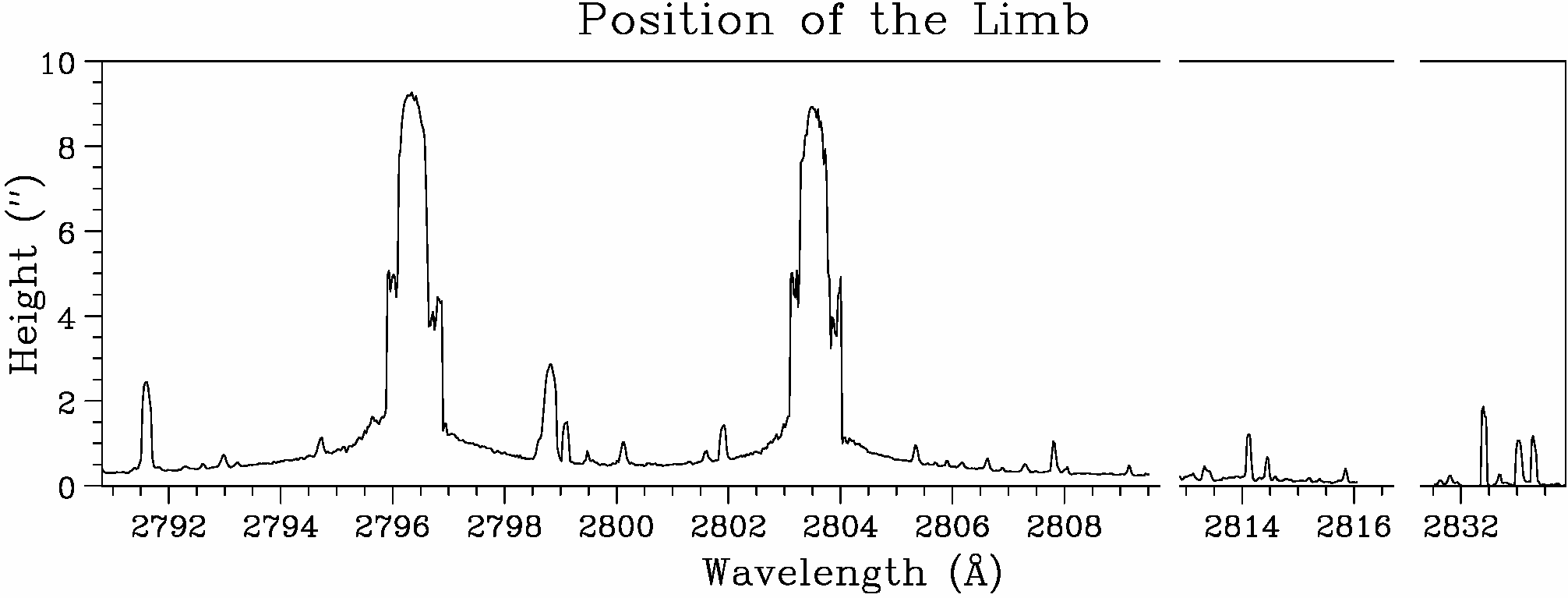}
  \caption{The height of the limb as a function of wavelength for the three spectral windows in the NUV spectral band of IRIS. The height was measured with respect to the limb position at 2832 \AA. From Alissandrakis et al. (2018), reprinted by permission from: Springer Nature, Solar Physics, \copyright 2018.}
  \label{fig:06}
\end{figure}

As mentioned in the introduction, Alissandrakis et al. (2018) measured the limb height for a number of spectral lines and continua from IRIS spectra, using as reference the position of the limb at 2832\,\AA. This is the wavelength of maximum intensity in the near ultraviolet (NUV) IRIS spectral band, located in the far red wing of the { k} line, but still below the local continuum level.  A plot of the height as a function of wavelength for the three spectral windows of IRIS NUV band is shown in Fig.~\ref{fig:06}; it was computed from averaged spectra, corrected for light scattered beyond the limb. 

\begin{figure}[!h]
  \centering
  \includegraphics[width=8.8cm]{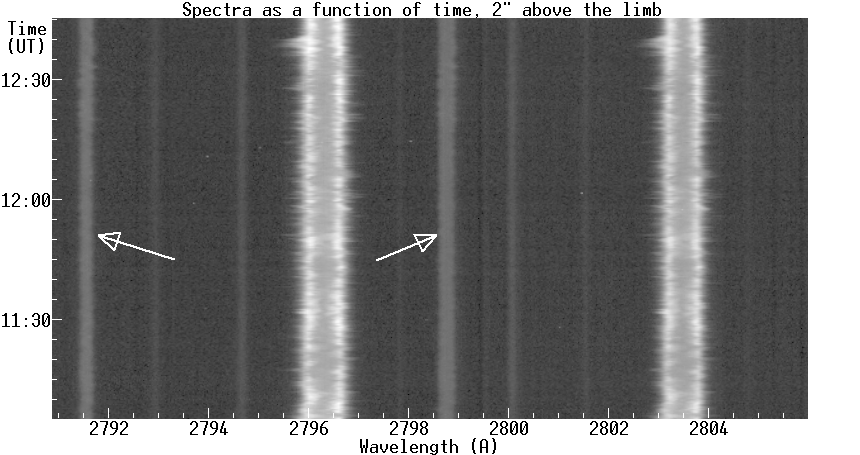}
  \caption{The intensity as a function of wavelength and time in the region of k and h lines, at a distance of 2\arcsec\ above the limb. The width of the cuts is 0.5\arcsec. The arrows point to the Mg{\sc\,ii} triplet lines. From Alissandrakis et al. (2018), reprinted by permission from: Springer Nature, Solar Physics, \copyright 2018.}
  \label{fig:triplet}
\end{figure}

In addition to the Mg{\sc\,ii} k and h lines at 2796.32 and 2803.58\,\AA\ respectively, many weaker lines extend above the height of the local continuum limb in this plot. Of particular interest are the Mg{\sc\,ii} triplet lines at 2791.59 and 2798.79\,\AA, not only because they extend more than 2\arcsec\ above the 2832\,\AA\ limb, but also because their variation with time does not reveal any spicular structure, which is very prominent in the k and h lines (Fig.~\ref{fig:triplet}). This may suggest the presence of an interspicular region in the chromosphere and/or that spicules are formed above the height of 2\arcsec, the chromosphere being more homogeneous up to that height.

\section{Variation of the network height with the solar cycle}\label{cycle}
In this section we investigate further a possible solar cycle variation of the network height, reported in paper II, based on the analysis of 6 observations. To this end, we expanded the data set by adding 7 more dates, so that we now cover the full interval of SDO observations from 2010 to the present, with about one observation per year.

\begin{figure}[h]
  \centering
  \includegraphics[width=8.8cm]{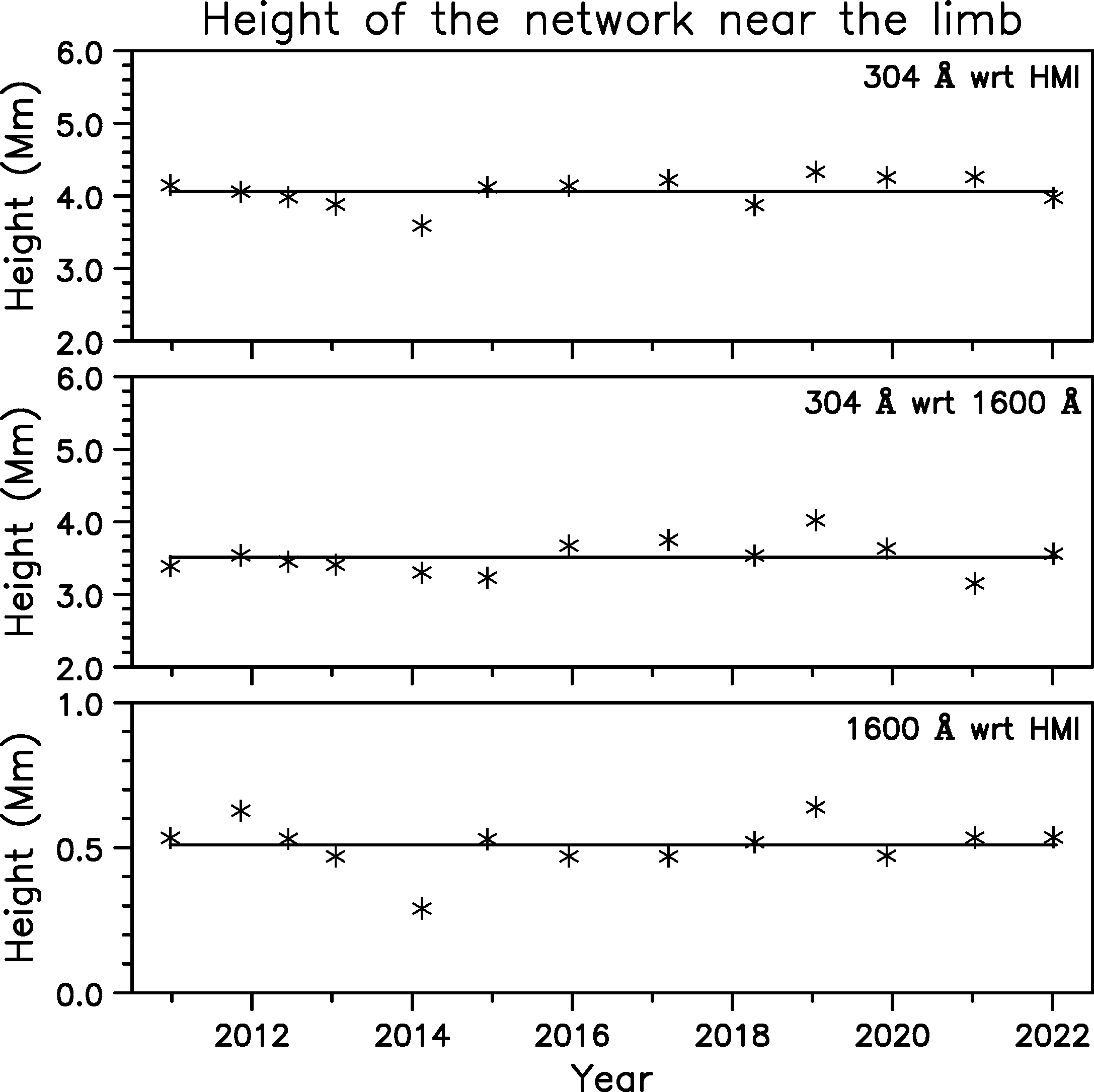}
  \caption{The height network structures as a function of time for a period slightly longer than a solar cycle, from SDO data. Horizontal lines mark the average value.}
  \label{fig:SolarCycle}
\end{figure}

Fig.~\ref{fig:SolarCycle} shows the network height as a function of time, extrapolated to the solar limb, for the 1600\,\AA\ and 304\.\AA\ network structures with respect to continuum magnetic features, as well as that of the 304\.\AA\ network with respect to 1600\,\AA. As in paper II, the values that were obtained by comparison to HMI magnetograms were corrected for the 0.34\,Mm shift between magnetic and intensity features near the limb. We note that variations are small, with no apparent association with the solar cycle.

\begin{table}[h]
\caption{Network formation height and limb position with respect to white light}
\label{Table:netheight}
\begin{center}
\begin{tabular}{cccc}
\hline
Wavelength&Disk center&Near the limb&Limb position\\
\AA       &Mm         &Mm           &Mm\\
\hline
1700 &$0.14\pm0.04$ &$0.24\pm0.11$&$0.72\pm0.04$ \\
1600 &$0.31\pm0.09$ &$0.50\pm0.08$&$1.13\pm0.25$ \\
~~304&$3.31\pm0.18$ &$4.06\pm0.20$&$9.25\pm2.62$ \\
\hline
\end{tabular}
\end{center}
\end{table}

As the extended data set gives more precise values for the network height, we present the new results in Table~\ref{Table:netheight}, where we have added a column with the position of the (outer) limb measured in paper I. We note that the latter is clearly above the level of the network near the limb, an effect that should be attributed to spicules. For reference, we note that Fossum \& Carlsson (2005) computed the formation height of the TRACE 1700 and 1600\,\AA\ wavelength bands at 0.36 and 0.43\,Mm, with widths of 0.325 and 0.185\,Mm respectively.

\section{Discussion and conclusions}\label{discuss}
The measurement of the position of the { CCTR} is based on the implicit assumption that the notion of the quiet sun (QS) is valid. We know very well that the quiet sun is not that quiet, with a multitude of small scale transient events and with structures like { spicules 
extending into the corona, 
having their}
own { CCTR}. Such phenomena can be considered as perturbations on top of the QS background; alternatively, the QS can be considered as the spatial, over the horizontal dimension, and temporal average of the actual physical conditions.

From the observational point of view, as measurements in different spectral bands are required, an important issue is the precision of the image scale. This effect is minimized when measurements are performed with the same instrument, e.g. with TRACE, IRIS or the AIA/HMI combination in SDO. It goes without saying that accurate pointing is required and that position measurements with respect to the WL limb are less susceptible to image scale errors than those with respect to the center of the solar disk.

Among the observations discussed in this work, the one that uses the position of the inner limb in EUV AIA images, developed in paper I, provides the most direct measurement of the height of the base of the transition region. The principal uncertainty of this approach is the correction for spicular absorption; without this correction we have an upper limit of 3.7\,Mm, which drops to 3\,Mm after the correction is applied{. This is clearly above the classical value of $\sim2$\,Mm, but is is well within the range predicted by rMHD models, such as that of Carlsson et al. (2016).} 

In order to put in the same framework the measurements of the formation height of the network, we should consider that the network emission is not formed at the same level as the average QS emission. Multi-component atmospheric models were first introduced by Vernazza et al. (1981), to describe specific atmospheric structures, such as the network and inter-network regions. Such models place the { CCTR} of each component at a different height, for example in the models of Fontenla et al. (1993) the base of the { CCTR} for the network is at 2.25\,Mm, while it is at 2.02\,Mm for the inter-network and 2.20\,Mm for the average QS. These differences are small and the formation height differences in various spectral bands are expected to be also small. Moreover, these models predict too much network/internetwork contrast at mm-$\lambda$, compared to ALMA results (Alissandrakis et al., 2020), which implies that network/internetwork differences are even smaller than the model predictions. We may thus consider that the network shift results are a fair approximation of the average QS formation height. In any case, we should bear in mind that the radiation forms in an extended height range, as shown in 
{ the left panel of Fig.~\ref{fig:models}}.

Direct measurements of the height of the solar limb in lines and continua that do not show spicular structure, such as the the Mg{\sc\,ii} triplet lines, place a lower limit to the height of the chromosphere. Still, as the region of formation of the radiation has a certain vertical extent, the position of the limb may represent the upper bound of this region rather than the average. In IRIS measurements, the uncertainty comes from the fact that the height is measured with respect to the quasi-continuum at 2832\,\AA, which should be formed slightly above the true continuum.

The situation is more complicated in lines and continua that show spicules. In such cases the height of the limb reflects the average height of the spicule forest rather than that of the homogeneous chromosphere. Measurements with ALMA fall in this group, where, in addition, the low spatial resolution of full-disk images limits the accuracy of the measurements.

\begin{figure}[t]
  \centering
  \includegraphics[width=8.8cm]{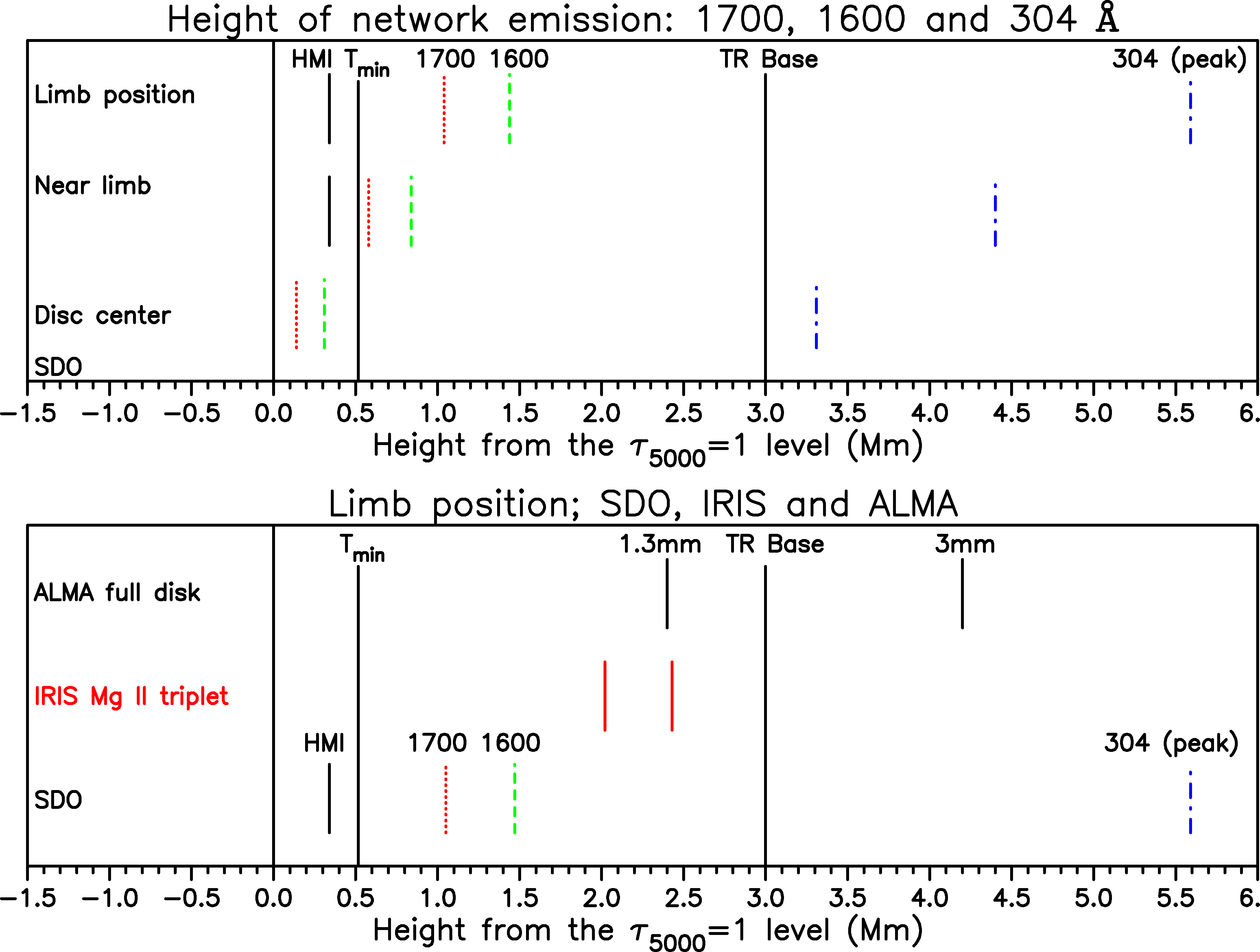}
  \caption{The height of the network emission (top) and the position of the limb (bottom) in Mm with respect to $\tau_{5000}=1$ level from various data sets discussed in this article. The position of the base of the { CCTR} from paper I, as well as that of the temperature minimum according to the model of Fontenla et al. (1993) are also marked.}
  \label{fig:cartoon}
\end{figure}

With all of the above considerations in mind, we have put together all the relevant observational results in Fig.~\ref{fig:cartoon}, where zero height is at the level of $\tau_{5000}=1$. The top panel shows the network height near disk center and near the limb from Table~\ref{Table:netheight}, assuming that the network forms at the same height as the average quiet sun and adding the 0.34\,Mm difference between the $\tau_{5000}=1$ and the limb to the network height near the limb. In the same plot we included the limb position from paper I, for reference; for 304\,\AA\ we put the height of the peak intensity, rather than that of the outer limb. As noted in Sect.~\ref{cycle}, the limb is well above the network, due to spicules. Another interesting remark is that, even at the center of the solar disk, the 304\,\AA\ network is above the base of the { CCTR}. { We note the AIA 304\,\AA\ channel is dominated by the He{\sc\,ii} lines which form around $T_e=50\,000$\,K, but this is very close to the top of the chromosphere (about 0.01\,Mm higher in the C7 model), as the CCRT is very steep in this temperature range.} This might imply that the { CCTR} is higher than estimated in paper I, but a more plausible explanation is that it is due to the effect of spicules, which have a very large optical depth at 304\,\AA, as evidenced in limb images where they extend much higher than their counterparts in other wavelengths (see, e.g. Fig. 9 in paper I). Their high optical depth makes them opaque on the disk, so that what we see there is not the network, but rather the base of spicules that are located on top of the network structures.

In the bottom panel of Fig.~\ref{fig:cartoon} we see that the Mg{\sc\,ii} triplet lines are comfortably below the base of the { CCTR}, and above the 1700/1600\,\AA\ limb. At mm-$\lambda$, the limb at 3\,mm is above, while at 1.3\,mm it is below the { CCTR}; we should note that, due to the low resolution of the full-disk images, the values have a large uncertainty: $\pm2.5$ and $\pm1.$4\,Mm respectively. We add that numerical simulations under local thermodynamic equilibrium by Mart\'inez-Sykora et al. (2020) gave average heights of 0.9\,Mm (with a standard deviation of 0.7\,Mm) for 1.2\,mm and 1.8\,Mm (standard deviation of 1\,Mm) at 3\,mm, whereas computations by the same authors under non-equilibrium hydrogen ionization gave greater heights and a much smaller height difference: $2.67\pm1.08$\,Mm for 1.2\,mm emission and $2.78\pm1.09$\,Mm for 3\,mm.

All things considered, the height of the base of the { CCTR} reported in paper I, $3.0\pm0.5$\,Mm, { which is consistent with the other observational results presented here, is definitely above the height predicted by classic homogeneous models and broadly consistent with the predictions of rMHD models}. Possible pole-equator differences, together with the effect of coronal holes, require further investigation. In paper I we found a pole/equator height ratio from { CCTR} images of just $1.08\pm0.02$, while we confirmed the prolate shape of the chromosphere, reported by Auch\`ere et al. (1998), in the 304\,\AA\ images; no pole/equator effect was found in the network height. As for solar cycle variations, our extended data set did not confirm the possible variation of the the network height reported in paper II. Concerning solar cycle variations of the height of the { CCTR}, the question is open.

\section{Acknowledgments}
The data used in his work were retrieved from the data bases of TRACE, SDO, IRIS and ALMA. The author is grateful to all those that worked for the development and operation of the instruments and made the data freely available to the community. { Thanks are also due to the reviewers for their comments that helped the author to improve the quality of this article.}

\section{References}
\small

\section*{References}
\small

\smallskip\noindent
Alissandrakis, C. E. (2019). Measurement of the Height of the Chromospheric 
Network Emission from Solar Dynamics Observatory Images (Paper II). Solar Phys., 
294(11), 161. doi:10.1007/s11207-019-1552-1. 381 arXiv:1911.00758.

\smallskip\noindent 
Alissandrakis, C. E., Nindos, A., Bastian, T. S. et al. (2020). Modeling the quiet
Sun cell and network emission with ALMA. Astron. Astrophys., 640, A57.
doi:10.1051/0004-6361/202038461. arXiv:2006.09886.

\smallskip\noindent
Alissandrakis, C. E., \& Valentino, A. (2019a). Erratum: Correction to: 
Structure of the Transition Region and the Low Corona from TRACE and SDO
Observations Near the Limb. Solar Phys., 294(10), 146. doi:10.1007/s11207-019-1545-0.

\smallskip\noindent
Alissandrakis, C. E., \& Valentino, A. (2019b). Structure of the Transition Region 
and the Low Corona from TRACE and SDO Observations Near the Limb (Paper I). 
Solar Phys., 294(7), 96. doi:10.1007/s11207-019-1486-7. arXiv:1906.09497.

\smallskip\noindent
Alissandrakis, C. E., Vial, J. C., Koukras, A. et al. (2018). IRIS Observations
of Spicules and Structures Near the Solar Limb. Solar Phys., 293(2), 20.
doi:10.1007/s11207-018-1242-4. arXiv:1801.02082.

\smallskip\noindent
Alissandrakis, C. E., Zachariadis, T., \& Gontikakis, C. (2005). Trace Observations 
of Solar Spicules Beyond the Limb in Ly$\alpha$ and Civ. In D. Danesy,
S. Poedts, A. de Groof, \& J. Andries (Eds.), The Dynamic Sun: Challenges
for Theory and Observations. volume 11 of ESA Special Publication. Published on 
CDROM, id. 54.1.

\smallskip\noindent
Anzer, U., \& Heinzel, P. (2005). On the Nature of Dark Extreme Ultraviolet
Structures Seen by SOHO/EIT and TRACE. Astrophys. J., 622(1), 714–721.
doi:10.1086/427817.

\smallskip\noindent
Athay, R. G. (1976). The solar chromosphere and corona: Quiet sun.
volume 53 of Astronomy Space Science Library Series. doi:10.1007/978-94-010-1715-2.

\smallskip\noindent
Auch\`ere, F., Boulade, S., Koutchmy, S. et al. (1998). The prolate solar 
chromosphere. Astron. Astrophys., 336, L57–L60.

\smallskip\noindent
Avrett, E., Tian, H., Landi, E. et al. (2015). Modeling the Chromosphere
of a Sunspot and the Quiet Sun. Astrophys. J., 811(2), 87. doi:10.1088/0004-637X/811/2/87.

\smallskip\noindent
Avrett, E. H., \& Loeser, R. (2008). Models of the Solar Chromosphere and
Transition Region from SUMER and HRTS Observations: Formation of the
Extreme-Ultraviolet Spectrum of Hydrogen, Carbon, and Oxygen. Astrophys. J. 
Suppl., 175(1), 229–276. doi:10.1086/523671.

\smallskip\noindent
Battaglia, M., Hudson, H. S., Hurford, G. J. et al. (2017). The Solar X
Ray Limb. Astrophys. J., 843(2), 123. doi:10.3847/1538-4357/aa76da.
arXiv:1705.11044.

\smallskip\noindent
Carlsson, M., Hansteen, V. H., Gudiksen, B. V. et al. (2016). A
publicly available simulation of an enhanced network region of the
Sun. Astron. Astrophys., 585, A4. doi:10.1051/0004-6361/201527226.
arXiv:1510.07581.

\smallskip\noindent
Daw, A., Deluca, E. E., \& Golub, L. (1995). Observations and Interpretation
of Soft X-Ray Limb Absorption Seen by the Normal Incidence X-Ray Telescope. 
Astrophys. J., 453, 929–944. doi:10.1086/176453.

\smallskip\noindent
de la Cruz Rodr\'iguez, J., Leenaarts, J., Danilovic, S. et al. (2019). STiC:
A multiatom non-LTE PRD inversion code for full-Stokes solar observations. 
Astron. Astrophys., 623, A74. doi:10.1051/0004-6361/201834464. arXiv:1810.08441.

\smallskip\noindent
Fontenla, J. M., Avrett, E. H., \& Loeser, R. (1990). Energy Balance in the
Solar Transition Region. I. Hydrostatic Thermal Models with Ambipolar
Diffusion. Astrophys. J., 355, 700–718. doi:10.1086/168803.

\smallskip\noindent
Fontenla, J.M., Avrett, E. H., \& Loeser, R. (1991). Energy Balance in the Solar
Transition Region. II. Effects of Pressure and Energy Input on Hydrostatic
Models. Astrophys. J., 377, 712–725. doi:10.1086/170399.

\smallskip\noindent
Fontenla, J. M., Avrett, E. H., \& Loeser, R. (1993). Energy Balance in the
Solar Transition Region. III. Helium Emission in Hydrostatic, Constant-Abundance 
Models with Diffusion. Astrophys. J., 406, 319–345. doi:10.1086/172443.

\smallskip\noindent
Fossum, A., \& Carlsson, M. (2005). Response Functions of the Ultraviolet
Filters of TRACE and the Detectability of High-Frequency Acoustic Waves.
Astrophys. J., 625(1), 556–562. doi:10.1086/429614.

\smallskip\noindent
Gudiksen, B. V., Carlsson, M., Hansteen, V. H. et al. (2011). The stellar
atmosphere simulation code Bifrost. Code description and validation.
Astron. Astrophys., 531, A154. doi:10.1051/0004-6361/201116520.
arXiv:1105.6306.

\smallskip\noindent
Mart\'inez-Sykora, J., De Pontieu, B., de la Cruz Rodriguez, J. et al. (2020). The
Formation Height of Millimeter-wavelength Emission in the Solar Chromosphere. 
Astrophys. J. Letters, 891(1), L8. doi:10.3847/2041-8213/ab75ac. 
arXiv:2001.10645.

\smallskip\noindent
Menezes, F., \& Valio, A. (2017). Solar Radius at Subterahertz Frequencies
and Its Relation to Solar Activity. Solar Phys., 292(12), 195. doi:10.1007/s11207-017-1216-y. 
arXiv:1712.06771.

\smallskip\noindent
Rozelot, J. P., Kosovichev, A., \& Kilcik, A. (2015). Solar Radius Variations:
An Inquisitive Wavelength Dependence. Astrophys. J., 812(2), 91. doi:10.1088/0004-637X/812/2/91.

\smallskip\noindent
Rutten, R. J. (2020). SolO campfires in SDO images. arXiv e-prints.
arXiv:2009.00376.

\smallskip\noindent
Thuillier, G., Claudel, J., Djafer, D. et al. (2011). The Shape of the Solar Limb:
Models and Observations. Solar Phys., 268(1), 125–149. doi:10.1007/s11207-010-9664-7.

\smallskip\noindent
Vernazza, J. E., Avrett, E. H., \& Loeser, R. (1976). Structure of the solar
chromosphere. II. The underlying photosphere and temperature-minimum
region. Astrophys. J. Suppl., 30, 1–60. doi:10.1086/190356.

\smallskip\noindent
Vernazza, J. E., Avrett, E. H., \& Loeser, R. (1981). Structure of the solar
chromosphere. III. Models of the EUV brightness components of the quiet
sun. Astrophys. J. Suppl., 45, 635–725. doi:10.1086/190731.

\smallskip\noindent
Zhang, J., White, S. M., \& Kundu, M. R. (1998). The Height Structure of the 
Solar Atmosphere from the Extreme-Ultraviolet Perspective. Astrophys. J. 
Letters, 504(2), L127–L130. doi:10.1086/311587.
arXiv:astro-ph/9807175

\smallskip\noindent
Zirin, H. (1966). The solar atmosphere. Waltham, Mass.: Blaisdell.


\end{document}